\documentclass[twocolumn]{aastex63}
\usepackage{lineno}

\usepackage[utf8]{inputenc}
\usepackage{amsmath}
\usepackage{amssymb}
\usepackage{graphicx}
\usepackage{epstopdf}
\usepackage{hyperref}
\usepackage{times}
\usepackage{verbatim, longtable}	
\usepackage{multirow} 
\usepackage{CJKutf8}
\usepackage[figuresright]{rotating}

\newcommand{\D}{\displaystyle}

\shorttitle{Fermi blazars}
\shortauthors{Chen et al.}

\begin{document}
\title{General Physical Properties of Fermi blazars}

\correspondingauthor{Yongyun Chen}
\email{ynkmcyy@yeah,net}

\correspondingauthor{Qiusheng Gu}
\email{qsgu@nju.edu.cn}

\author{Yongyun Chen$^{*}$ \begin{CJK*}{UTF8}{gkai}(陈永云)\end{CJK*}}
\affiliation{College of Physics and Electronic Engineering, Qujing Normal University, Qujing 655011, P.R. China}


\author{Qiusheng Gu$^{*}$ \begin{CJK*}{UTF8}{gkai}(顾秋生)\end{CJK*}}
\affiliation{School of Astronomy and Space Science, Nanjing University, Nanjing 210093, P. R. China}


\author{Junhui Fan \begin{CJK*}{UTF8}{gkai}(樊军辉)\end{CJK*}}
\affiliation{Center for Astrophysics,Guang zhou University,Guang zhou510006, China}

\author{Xiaoling Yu\begin{CJK*}{UTF8}{gkai}(俞效龄)\end{CJK*}}
\affiliation{College of Physics and Electronic Engineering, Qujing Normal University, Qujing 655011, P.R. China}

\author{Xiaogu Zhong\begin{CJK*}{UTF8}{gkai}(钟晓谷)\end{CJK*}}
\affiliation{College of Physics and Electronic Engineering, Qujing Normal University, Qujing 655011, P.R. China}

\author{Hongyu Liu\begin{CJK*}{UTF8}{gkai}(刘红宇)\end{CJK*}}
\affiliation{College of Physics and Electronic Engineering, Qujing Normal University, Qujing 655011, P.R. China}

\author{Nan Ding \begin{CJK*}{UTF8}{gkai}(丁楠)\end{CJK*}}
\affiliation{School of Physical Science and Technology, Kunming University 650214, P. R. China}

\author{Dingrong Xiong \begin{CJK*}{UTF8}{gkai}(熊定荣)\end{CJK*}}
\affiliation{Yunnan Observatories, Chinese Academy of Sciences, Kunming 650011,China}

\author{Xiaotong Guo \begin{CJK*}{UTF8}{gkai}(郭晓通)\end{CJK*}}
\affiliation{Anqing Normal University, 246133, P. R. China}

\begin{abstract}
We study the general physical properties of Fermi blazars using the Fermi fourth source catalog data (4FGL-DR2). The quasi-simultaneous multiwavelength data of Fermi blazar are fitted by using the one-zone leptonic model to obtain some physical parameters, such as jet power, magnetic field and Doppler factor. We study the distributions of the derived physical parameter as a function of black hole mass and accretion disk luminosity. The main results are as follows. (1) For a standard thin accretion disk, the jet kinetic power of most FSRQs can be explained by the BP mechanism. However, the jet kinetic power of most BL Lacs can not be explained by both the BZ mechanism or the BP mechanism. The BL Lacs may have ADAFs surrounding their massive black holes. (2) After excluding the redshift, there is a moderately strong correlation between the jet kinetic power and jet radiation power and the accretion disk luminosity for Fermi blazars. These results confirm a close connection between jet and accretion. The jet kinetic power is slightly larger than the accretion disk luminosity for Fermi blazars. (3) There is a significant correlation between jet kinetic power and gamma-ray luminosity and radio luminosity for Fermi blazars, which suggests that gamma-ray luminosity and radio luminosity can be used to indicate the jet kinetic power.
\end{abstract}


\section{Introduction}
The blazar is a special subclass of active galactic nuclei (AGN), whose jet is directed at the observer \citep{Urry1995}. Blazar are usually divided into flat spectrum radio quasars (FSRQs) and BL Lac objects (BL Lacs) according to the equivalent width of optical emission lines. The FSRQs have strong broad emission lines, and the EW of their emission lines is greater than 5\AA. The BL Lacs show weak or no broad emission lines, and its EW of emission lines is less than 5\AA. However, arbitrary classification based on EW is not enough. On one hand, \cite{Blandford1978} suggested that the lack of broad emission lines in the BL Lacs may be due to the Doppler-boosted continuum swamping out any spectral lines. On the other hand, the detection of broad emission lines may be the result of the low jet activity states \citep{Vermeulen1995}. Therefore, the physical difference between FSRQs and BL Lacs can not be revealed by the blazar classification mechanism of EW. Some authors have proposed a more physical classification mechanism for blazars. \cite{Ghisellini2011} and \cite{Sbarrato2012} proposed that the broad-line region (BLR) luminosity ($L_{\rm BLR}$) in Eddington units can distinguish FSRQs from BL Lacs. The FSRQs have $L_{\rm BLR}/L_{\rm Edd}\geq10^{-3}$ or $L_{\rm BLR}/L_{\rm Edd}\geq5\times10^{-4}$,  which, in turn, indicates a
radiatively efficient accretion process ($L_{\rm disk}/L_{\rm Edd}\geq0.01$). The BL Lacs have a radiatively inefficient accretion flow. 

The formation of relativistic jets in AGN has always been a hot issue in astrophysics, and its formation mechanism has not been clear. Many theoretical models have been proposed to explain the formation of the jets. Among the current theoretical models of jet formation, there are two most popular theories. One is the Blandford-Znajek (BZ) mechanism \citep{Blandford1977}, in which the jet extracts the rotational energy of the black hole. The other is the Blandford-Payne (BP) mechanism \citep{Blandford1982}: the jets mainly extract the rotational energy of the accretion disk. In both cases, the magnetic field plays a major role in directing power from a black hole (BH) or disk into the jet; in both cases, it should be maintained by matter accreted to the black hole, leading to an expected relationship between accretion and jet power \citep{Maraschi2003}. \cite{Chai2012} found a significant correlation between black hole mass and the bulk Lorentz factor using 101 radio-loud AGNs. They suggested that the BZ mechanism may dominate over the BP mechanism in these radio-loud AGNs. \cite{Zhang2022} proposed that the jets of both FSRQs and BL Lacs are likely produced by the BZ mechanism. \cite{Foschini2011} compared the maximum jet power of the BZ mechanism with the observed jet power and suggested that the jets of FSRQs can not be fully explained by the BZ mechanism. However, at present, it is not clear whether the BZ mechanism or the BP mechanism dominates the jet formation of the Fermi blazars.        

Since the successful launch of the Fermi Space Telescope, many AGNs have been detected with high-energy gamma-ray radiation \citep{Abdo2009, Abdo2010, Nolan2012, Acero2015, Abdollahi2020, Ajello2022}, especially the blazars, confirming that these AGNs have strong relativistic jets. Encouraged by the availability of high-quality multi-wavelength (MW) data sets of large samples of Fermi blazars, we systematically study their broadband properties by using observational and theoretical SED modeling methods. Our main goal is to study the basic properties of Fermi blazars, such as the relationship between accretion and jets, and the formation mechanism of the jets. Compared with previous studies, we focus on the observed results and then use a leptonic emission model to explain them. Here, we show the results of our research on blazars included in the Fermi Large Area Telescope (LAT)
fourth source catalog data release 2 (4FGL-DR2; \cite{Abdollahi2020}). Our sample is probably the largest sample that applies the physical SED model. In Section 2, we describe the sample. Section 3 shows the model of jets. Section 4 describes the results and discussion. Section 5 is the conclusion.

\section{The sample}
\subsection{The Fermi blazars sample}
\cite{Paliya2021} used the 4FGL-DR2 catalog to get the 1077 Fermi blazars with black hole mass and accretion disk luminosity. We carefully examined the sample of \cite{Paliya2021} and compared it with the source classification of \cite{Abdollahi2020} and \cite{Foschini2021}. We only consider sources that has reliable redshift, black hole mass, accretion disk luminosity, 1.4 GHz radio flux, and quasi-simultaneous multiwavelength data. The redshift, black hole mass, and accretion disk luminosity come from the work of \cite{Paliya2021}. The 1.4 GHz radio flux comes from the NASA/IPAC Extragalactic Database. The quasi-simultaneous multiwavelength data comes from the Space Science Data Center (SSDC) SED Builder \footnote{http://tools.ssdc.asi.it/SED/}. We get 459 Fermi blazars (317 FSRQs, and 142 BL Lacs: 41 low synchrotrons peaked BL Lacs (LBLs), 18 intermediate synchrotrons peaked BL Lacs (IBLs), and 83 high synchrotrons peaked BL Lacs (HBLs)). The boundaries of BL Lacs are $\log\nu_{\rm p}<14$ Hz for LBLs, Hz $14<\log\nu_{\rm p}<15$ Hz for IBLs, and $\log\nu_{\rm p}>15$ Hz for HBLs \citep[e.g.,][]{Abdo2010b}, respectively. 

\subsection{The jet power}
\begin{figure}
	\includegraphics[width=8.5cm,height=8.5cm]{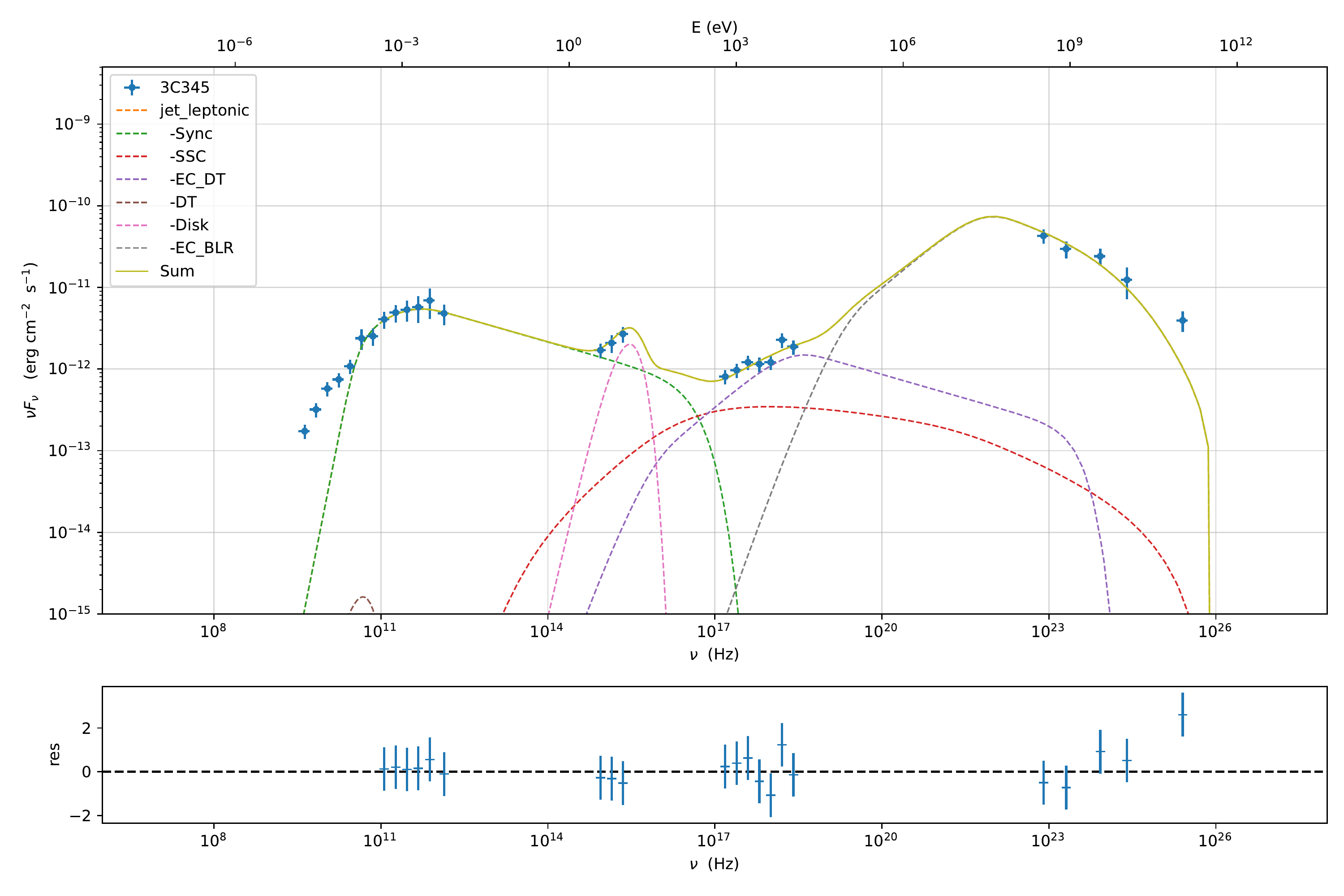}
	\caption{The example of broadband SEDs of 3C345 is modeled by using a one-zone model.}
	\label{Figure1}
\end{figure}

The quasi-simultaneous multiwavelength data of Fermi blazars are modeled with a
simple one-zone leptonic emission model \citep{Tramacere2009, Tramacere2011, Tramacere2020}. The broadband SEDs have been modeled using
the open source package JetSet\footnote{https://jetset.readthedocs.io/en/latest/
} numerical leptonic code \citep{Tramacere2020}. According to the the minimum $\chi^{2}/dof$, the parameters were defined as the best fit values (see Figure~\ref{Figure1}). We estimate the jet power of electrons ($P_{\rm ele}$), Poynting flux ($P_{\rm mag}$), radiation ($P_{\rm rad}$), and protons ($P_{\rm p}$) as follows 

\begin{equation}
	P_{\rm i}=\pi R^{2}\Gamma^{2}\beta cU_{\rm i}^{'}
\end{equation}   
where $U_{\rm i}^{'}$ is the energy density of the i component, which is protons (i=p), relativistic electrons (i=e), the magnetic field (i=B) and the produced radiation (i=rad). The radiative power is derived as

\begin{equation}
P_{\rm rad}=\pi R^{2}\Gamma^{2}\beta cU_{\rm rad}^{'}
\end{equation}
where $U_{\rm rad}^{'}$ is the radiation energy density ($U_{\rm rad}^{'}=L^{'}/(4\pi R^{2}c)$). The $L^{'}$ is the total observed non-thermal luminosity in the comoving frame. The $\delta$ is the Doppler factor, $\delta=(\Gamma(1-\beta cos\theta))^{-1}$, where $\theta$ is the angle between the jet axis and line of sight of the observer. The $\Gamma$ is Lorentz factor.  For blazars, we have sin($\theta$)$\approx1/\Gamma$ and, thus, $\Gamma \simeq \delta$ \citep{Ghisellini2014}. The $\beta$ is the jet relativistic speed, $\beta=\sqrt{1-1/\Gamma^{2}}$. The size of the emission region can be derived
from the relation $R=ct_{var}\delta/(1+z)$ \citep{Ghisellini2014}, where $t_{var}$ is the variability timescale. The relevant data is shown in Table 1. The example is shown in Figure~\ref{Figure1}. 
    
\begin{longrotatetable}
	\begin{deluxetable*}{lllllllllllllllll}
		\tablecaption{The sample of Fermi blazars\label{chartable}}
		\tablewidth{700pt}
		\tabletypesize{\scriptsize}
		\tablehead{
			\colhead{Name} & \colhead{RA} & 
			\colhead{DEC} & \colhead{Type} & 
			\colhead{Redshift} & \colhead{$\log M_{\rm BH}$} & 
			\colhead{$L_{\rm disk}$} & \colhead{$\log L_{\gamma}$} & 
			\colhead{$\log L_{\rm radio}$} & \colhead{$B$} & \colhead{$\delta$} & \colhead{$\log P_{\rm rad}$} & \colhead{$\log P_{\rm e}$} & \colhead{$\log P_{\rm B}$} & \colhead{$\log P_{\rm p}$} & \colhead{$\log L_{\rm sy}$} &  \colhead{$\log \nu_{\rm syn}$}\\ 
			\colhead{(1)} & \colhead{(2)} & \colhead{(3)} & \colhead{(4)} & 
			\colhead{(5)} & \colhead{(6)} & \colhead{(7)} &
			\colhead{(8)} & \colhead{(9)} & \colhead{10} & \colhead{(11)} & \colhead{(12)} & \colhead{(13)} & \colhead{(14)} & \colhead{(15)} & \colhead{(16)} & \colhead{(17)}
		} 
		\startdata
J0003.2+2207	&	0.8058	&	22.1302	&	BLL	&	0.1	&	8.1	&	42.74	&	43.71	&	39.96	&	0.0104	&	6.82	&	44.48	&	45.54	&	43.92	&	47.26	&	43.23	&	15.15	\\
J0004.4-4737	&	1.1091	&	-47.6233	&	FSRQ	&	0.88	&	8.28	&	45.1	&	47.16	&	44.53	&	0.55	&	37.03	&	43.48	&	44.82	&	45.72	&	46.22	&	46.61	&	13.01	\\
J0006.3-0620	&	1.5992	&	-6.3493	&	BLL	&	0.347	&	8.92	&	44.52	&	45.01	&	43.62	&	4.85	&	2.53	&	45.73	&	45.29	&	46.72	&	46.89	&	45.74	&	12.92	\\
J0010.6+2043	&	2.6502	&	20.7332	&	FSRQ	&	0.598	&	7.86	&	45.34	&	45.94	&	43.19	&	0.0011	&	1.12	&	46.66	&	47.65	&	38.98	&	47.9	&	45.47	&	12.42	\\
J0016.2-0016	&	4.061	&	-0.2806	&	FSRQ	&	1.577	&	8.52	&	45.77	&	48.73	&	45.61	&	0.34	&	21.5	&	44.09	&	42.49	&	45.71	&	41.82	&	47.22	&	12.44	\\
J0019.6+7327	&	4.9031	&	73.456	&	FSRQ	&	1.781	&	9.31	&	46.62	&	49.32	&	45.99	&	0.037	&	39.99	&	45.26	&	46.22	&	43.99	&	45.89	&	47.77	&	12.29	\\
J0022.0+0006	&	5.5154	&	0.1134	&	BLL	&	0.306	&	8.02	&	43.79	&	44.9	&	40.79	&	0.121	&	22.32	&	47.29	&	44.9	&	44.74	&	45.73	&	44.57	&	16.67	\\
J0023.7+4457	&	5.9477	&	44.951	&	FSRQ	&	1.062	&	7.71	&	45.09	&	47.49	&	44.02	&	0.0071	&	142.42	&	42.13	&	46.87	&	42.46	&	47.34	&	46.56	&	12.8	\\
J0024.7+0349	&	6.1975	&	3.8321	&	FSRQ	&	0.546	&	7.11	&	44.62	&	45.91	&	42.21	&	0.134	&	25.58	&	42.86	&	42.57	&	46.7	&	43.24	&	45.15	&	13.47	\\
J0032.4-2849	&	8.1076	&	-28.8224	&	BLL	&	0.324	&	8.47	&	44.02	&	45.16	&	42.44	&	0.357	&	30.84	&	44.53	&	42.38	&	47.02	&	42.33	&	44.95	&	13.78	\\
J0038.2-2459	&	9.5652	&	-24.9899	&	FSRQ	&	0.498	&	8.14	&	44.97	&	45.92	&	43.36	&	0.234	&	14.15	&	43.82	&	44.18	&	46.4	&	45.23	&	45.59	&	12.43	\\
J0039.0-0946	&	9.7556	&	-9.7828	&	FSRQ	&	2.106	&	8.5	&	45.73	&	49.19	&	45.47	&	0.188	&	40.34	&	42.96	&	44.39	&	46.69	&	46.11	&	47.59	&	12.53	\\
J0042.2+2319	&	10.5581	&	23.3271	&	FSRQ	&	1.425	&	8.73	&	45.49	&	48.02	&	45.52	&	0.149	&	26.59	&	44.08	&	44.59	&	45.29	&	44.69	&	46.92	&	12.27	\\
J0043.8+3425	&	10.9717	&	34.4316	&	FSRQ	&	0.969	&	7.83	&	44.76	&	47.81	&	43.68	&	0.0139	&	69.65	&	43.57	&	46.75	&	41.62	&	46.02	&	45.89	&	13.77	\\
J0044.2-8424	&	11.0711	&	-84.4016	&	FSRQ	&	1.032	&	8.52	&	45.87	&	47.3	&	44.54	&	8.17	&	126.7	&	42.5	&	41.89	&	49.16	&	43.5	&	46.58	&	12.96	\\
J0045.1-3706	&	11.2936	&	-37.1065	&	FSRQ	&	1.015	&	8.61	&	45.86	&	47.51	&	44.31	&	0.073	&	23.77	&	43.8	&	45.85	&	44.92	&	47.3	&	46.73	&	12.39	\\
J0045.7+1217	&	11.4309	&	12.292	&	BLL	&	0.255	&	8.82	&	44.18	&	45.49	&	41.98	&	0.134	&	26.45	&	42.47	&	42.89	&	45.69	&	42.93	&	44.78	&	15.56	\\
J0047.9+2233	&	11.9981	&	22.5632	&	FSRQ	&	1.163	&	8.07	&	45.57	&	47.8	&	43.95	&	0.087	&	26	&	43.65	&	43.78	&	45.67	&	43.39	&	46.24	&	12.8	\\
J0049.6-4500	&	12.4188	&	-45.0086	&	FSRQ	&	0.121	&	8.08	&	43.51	&	44.03	&	41.52	&	0.46	&	39.4	&	45.43	&	43.57	&	47.23	&	43.98	&	44.39	&	12.55	\\
J0050.0-5736	&	12.5197	&	-57.6164	&	FSRQ	&	1.797	&	9.06	&	46.88	&	48.77	&	46.24	&	3.049	&	22.97	&	44.63	&	43.68	&	48.21	&	44.66	&	48	&	12.3	\\
J0051.1-0648	&	12.7824	&	-6.8096	&	FSRQ	&	1.975	&	9.31	&	47.11	&	49.12	&	46.08	&	0.0375	&	27.51	&	44.71	&	46.9	&	45.1	&	47.8	&	48.23	&	12.48	\\
J0056.3-0935	&	14.0874	&	-9.5997	&	BLL	&	0.103	&	8.96	&	43.22	&	44.39	&	41.36	&	0.096	&	35.19	&	47.28	&	47.2	&	44.81	&	47.41	&	43.92	&	15.78	\\
J0058.0-0539	&	14.5108	&	-5.655	&	FSRQ	&	1.246	&	8.7	&	46.28	&	47.86	&	45.03	&	0.259	&	22.41	&	44.02	&	44.64	&	46.74	&	45.79	&	46.81	&	12.56	\\
J0059.2+0006	&	14.8073	&	0.1166	&	FSRQ	&	0.719	&	8.56	&	46.08	&	46.34	&	44.63	&	0.213	&	27.23	&	43.95	&	44.14	&	46.67	&	43.77	&	46.12	&	13.17	\\
J0059.3-0152	&	14.8361	&	-1.8725	&	BLL	&	0.144	&	8.63	&	43.52	&	44.23	&	40.63	&	0.245	&	52.13	&	45.52	&	43.53	&	44.86	&	42.34	&	44.15	&	16.67	\\
J0102.8+5824	&	15.701	&	58.4092	&	FSRQ	&	0.644	&	9.01	&	46.04	&	47.44	&	44.02	&	0.153	&	27.03	&	45.54	&	45.52	&	45.95	&	45.71	&	46.57	&	12.86	\\
J0104.8-2416	&	16.2146	&	-24.2808	&	FSRQ	&	1.747	&	8.98	&	46.05	&	48.77	&	45.22	&	0.193	&	14.89	&	46.17	&	44.9	&	45.41	&	44.32	&	47.83	&	12.2	\\
J0105.1+3929	&	16.2913	&	39.4963	&	BLL	&	0.44	&	8.17	&	44.34	&	46.01	&	42.55	&	0.00948	&	82.07	&	42.47	&	46.33	&	41.37	&	45.95	&	45.76	&	13.45	\\
		\enddata
		\tablecomments{Columns (1) is the 4FGL name of sources; Columns (2) is the Right ascension in decimal degrees; Columns (3) is Declination in decimal degrees; Columns (4) is the Class of sources; Columns (5) is the redshift; Columns (6) is the black hole mass; Columns (7) is the accretion disk luminosity (erg s$^{-1}$); Columns (8) is the gamma-ray luminosity (erg s$^{-1}$); Columns (9) is the 1.4 GHz radio luminosity (erg s$^{-1}$); Columns (10) is the magnetic field (Gauss); Columns (11) is the Doppler factor; Columns (12) is the radiation jet power (erg s$^{-1}$); Columns (13) is the electron jet power (erg$~s^{-1}$); Columns (14) is the magnetic field jet power (erg s$^{-1}$); Columns (15) is the protons jet power (erg s$^{-1}$); Columns (16) is the synchrotron peak frequency luminosity (erg s$^{-1}$); Columns (17) is the synchrotron peak frequency. This table is available in its entirety in machine-readable form.}
	\end{deluxetable*}
\end{longrotatetable}

\section{The jet model} 
Some authors have calculated the maximum jet power that can be extracted from a rapidly rotating black hole / magnetized accretion disk \citep{Ghosh1997, Livio1999, Cao2003}, namely the BZ and BP mechanisms. Our calculation mainly follows their method.

\subsection{The BP model}
The jet power of the maximum BP model can be calculated by the following formula

\begin{equation}
	P_{\rm BP}^{\rm max}=4\pi\int\limits\frac{B_{\rm pd}^{2}}{4\pi}R^{2}\Omega(R)dR
\end{equation}
where $B_{\rm pd}$ is the strength of a large-scale ordered field on the surface of a disk. \cite{Livio1999} showed that the large-scale magnetic field threading the disk and the magnetic field generated by the  dynamo processes are approximately

\begin{equation}
	B_{\rm pd}\sim \frac{H}{R}B_{\rm dynamo}.
\end{equation}

The scale height of the disk ($H/R$) is estimated as follows \citep{Laor1989}

\begin{equation}
	\frac{H}{R} = 15.0\dot{m}r^{-1}c_{2}, 
\end{equation}
where the coefficient $c_{2}$ is defined by \cite{Novikov1973}, and other
parameters are defined by

\begin{eqnarray}
	r = \frac{R}{R_{\rm G}}, R_{\rm G} =\frac{GM_{\rm bh}}{c^{2}}, \dot{m} = \frac{\dot{M}}{\dot{M_{\rm Edd}}}, \nonumber \\
	\dot{M_{\rm Edd}} = \frac{L_{\rm Edd}}{\eta_{\rm eff}c^{2}}=1.39\times10^{15} m~kg~s^{-1}, m = \frac{M_{\rm bh}}{M_{\odot}}	
\end{eqnarray}
where $\eta_{\rm eff}=0.1$ is adopted. 

\cite{Cao2003} has given the expression of the magnetic field produced by
dynamo processes in the disk as

\begin{equation}
	B_{\rm dynamo} = 3.56\times10^{8}r^{-3/4}m^{-1/2}A^{-1}BE^{1/2} G.
\end{equation}
\cite{Novikov1973} has defined the general relativistic correction factors $A,B,$ and $E$. The $A,B$,and $E$ are estimated by the following formula,

\begin{eqnarray}
	A=1+a^{2}x^{-4}+2a^{2}x^{-6}, \nonumber \\
	B=1+ax^{-3}, \nonumber \\
	E=1+4a^{2}x^{-4}-4a^{2}x^{-6}+3a^{4}x^{-8}, \nonumber \\
	x = \sqrt r
\end{eqnarray} 

The Kepler angular velocity for standard accretion disk models is 

\begin{equation}
	\Omega(r) = \frac{2.03\times10^{5}}{m(a+r^{3/2})} s^{-1},	
\end{equation}
where $a$ is the spin of black hole \citep{Cao2003}. 

According to equations (3) to (9), the maximal jet power of the BP model can be obtained, if some parameters ($m$, $\dot{m}$, $a$) are specified. 

\subsection{The BZ model} 
\cite{Livio1999} suggested that the jet power of the BZ mechanism is determined
by the hole mass ($m$), the spin of the black hole ($a$), and the strength
of the poloidal field threading the horizon of the black hole. The maximum jet power of the BZ model can be estimated by the following formula \citep[e.g.][]{Ghosh1997, MacDonald1982},

\begin{equation}
	P_{\rm BZ}^{\rm max} = \frac{1}{32}\omega_{F}^{2}B_{\perp}^{2}R_{\rm H}^{2}ca^{2},
\end{equation}
where $R_{\rm H}$ is the the horizon radius, $R_{\rm H}=[1+(1-a^{2})^{1/2}]GM_{\rm bh}/c^{2}$. We use $\omega_{F} = 1/2$, $B_{\perp}\simeq B_{\rm pd}(R_{\rm ms})$ \citep[e.g.][]{MacDonald1982, Livio1999, Cao2003} to estimate the maximum jet power of the BZ model. The $R_{\rm ms}$ is defined by the following formula,

\begin{eqnarray}
	R_{\rm ms} = R_{\rm G}\{3+Z_{2}-\left[(3-Z_{1})(3+Z_{1}+2Z_{2})\right]^{1/2}\},\nonumber \\
	Z_{1}\equiv1+(1-a^{2})^{1/3}\left[(1+a)^{1/3}+(1-a)^{1/3}\right], \nonumber \\
	Z_{2}\equiv(3a^{2}+Z_{1}^{2})^{1/2}.
\end{eqnarray}  
We use the equations (4), (7), and (10) to calculate the maximal jet power of the BZ model. The scale height of the disk is proportional to the dimensionless accretion rate $\dot{m}$. Thus, the maximal jet power extracted from the disk or the spinning black hole depends on the accretion rate $\dot{m}$. In this work, the accretion rate $\dot{m}$ is adopted as a free parameter to compare our calculations with observations. The spin of black hole $a=0.95$ is adopted  \citep{Cao2002b, Cao2003}. 

\section{Results and Discussion}
\begin{figure}
    \includegraphics[width=8.5cm,height=17cm]{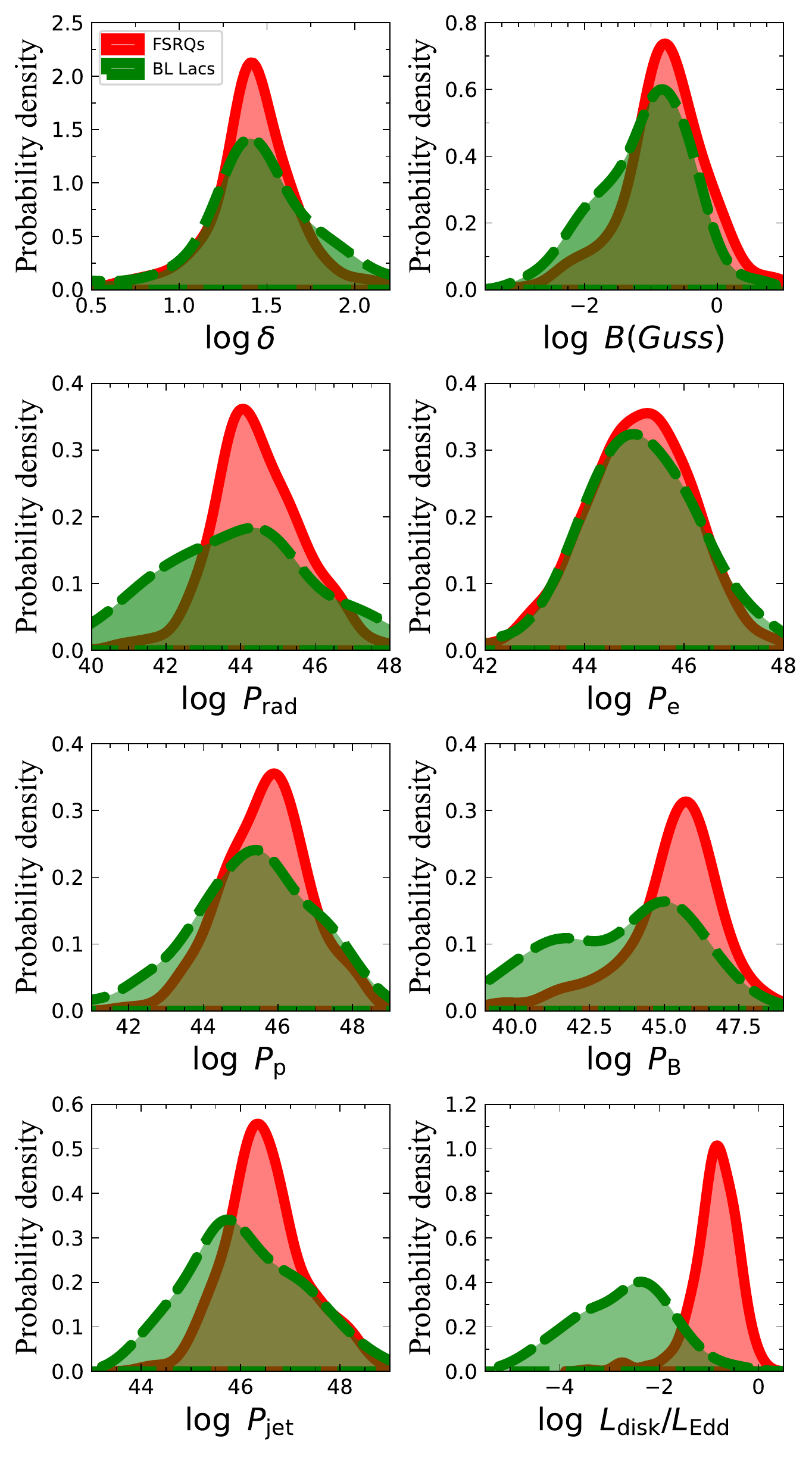}
    \caption{The distribution of physical parameters. The jet power of kinetic is $P_{\rm jet}=P_{\rm B}+P_{\rm e}+P_{\rm p}$. The red line is FSRQs and the green line is BL Lacs.}
    \label{Figure2}
\end{figure}

The distribution of the physical parameters of the jets of Fermi blazars is shown in Figure~\ref{Figure2}. The red-shaded areas are FSRQs, and the green-shaded areas are BL Lacs. We study the difference in the physical parameter distribution of jets by using a parametric T-test, nonparametric Kruskal-Wallis H Test, and nonparametric Kolmogorov–Smirnov (K-S) test. The parameter T-test is mainly used to test whether there is a difference in the average value of two independent samples. The nonparametric Kruskal-Wallis H Test and nonparametric Kolmogorov–Smirnov (K-S) test are mainly used to test whether there are differences in the distribution of physical parameters between two independent samples. We assume that there are differences in the distribution of two independent samples among three tests simultaneously.

\subsection{The Doppler factor}
\begin{figure}
	\includegraphics[width=8.5cm,height=8.5cm]{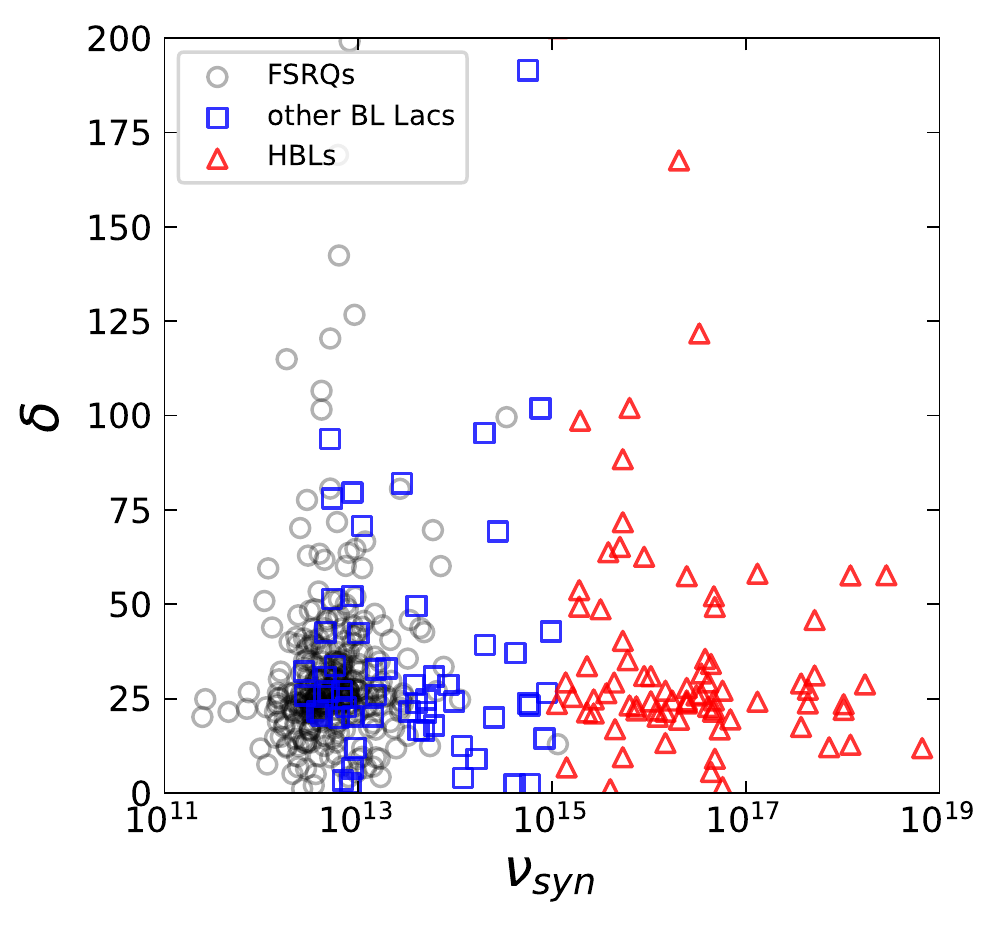}
	\caption{The Doppler factor versus synchrotron peak frequency for Fermi blazars. The black circle is FSRQs; the blue square is other BL Lacs; the red triangle is HBLs.}
	\label{Figure0}
\end{figure}

The average Doppler factors of FSRQs and BL Lacs are $\langle\log\delta_{\rm FSRQs}\rangle=1.43$ and $\langle\log\delta_{\rm BL Lacs}\rangle=1.45$, respectively.
According to the T-test ($P=0.45$, significant probability $P<0.05$), Kolmogorov–Smirnov test ($P=0.04$, significant probability $P<0.05$), and Kruskal-Wallis H Test ($P=0.51$, significant probability $P<0.05$), we find that the distributions of Doppler factor between FSRQs and BL Lacs are not significantly different. \cite{Weaver2022} found a higher probability for FSRQs to have a larger Doppler factor than BL Lacs by using the Very Long Baseline Array (VLBA) at 43 GHz. 
\cite{Liodakis2018} also found that the FSRQs ($\delta_{var}=11$) have a slightly higher Doppler factor than BL Lacs ($\delta_{var}=10$) using 1029 sources observed by the Owens Valley Radio Observatory’s 40 m telescope. However, we find that the FSRQs have a slightly lower Doppler factor than BL Lacs. \cite{Lister2019} studied the parsec-scale jet kinematics of 409 bright radio-loud blazars based on 15 GHz Very Long Baseline Array (VLBA). \cite{Lister2019} found that the AGN with low synchrotron peak frequencies has the highest jet speeds (namely high Doppler factor). Figure~\ref{Figure0} shows the relation bteween Doppler factor and synchrotron peak frequency for Fermi blazars. We find that the AGN with high synchrotron peak frequencies tends to have high Doppler factor. Our results are slightly contradictory to the results of \cite{Lister2019}. There is a possible explanation that the contradiction between our results and previous studies is that there are different methods for calculating Doppler factors. \cite{Liodakis2015} also found that the Doppler factors obtained by different methods were different. \cite{Weaver2022} also found that the distributions of the Doppler factor between FSRQs and BL Lacs are not significantly different based on the K-S test ($P=0.333$). Our results are consistent with the results of \cite{Weaver2022}. According to the broadband spectral energy distributions (SEDs), \cite{Chen2018} found that the mean values of the Doppler factor for FSRQs and BL Lacs are 13.87 and 27.33, respectively. The BL Lacs have a higher average Doppler factor than FSRQs. Our results are consistent with the results of \cite{Chen2018}.     
  
\subsection{The Magnetic field} 
The average values of magnetic field of FSRQs and BL Lacs are $\langle\log B_{\rm FSRQs}\rangle=-0.79$ and $\langle\log B_{\rm BL Lacs}\rangle=-1.09$, respectively.   
There is a significant difference in the average magnetic field between FSRQs and BL Lacs using the T-test ($P=7.97\times10^{-6}$). Through a Kolmogorov–Smirnov test ($P=0.002$) and Kruskal-Wallis H Test ($P=3.38\times10^{-5}$), we also find that the distributions of magnetic field between FSRQs and BL Lacs are significantly different. The FSRQs have a higher average magnetic field than BL Lacs. \cite{Ghisellini2010} also found that FSRQs have a strong magnetic field than BL Lacs by using 89 Fermi blazars. This is also in agreement with findings from radio VLBA observations \citep{Pushkarev2012}. 
   
\subsection{The jet power}
The average values of jet powers of radiation of FSRQs and BL Lacs are $\langle\log P_{\rm rad, FSRQs}\rangle=44.48$ and $\langle\log P_{\rm rad, BL Lacs}\rangle=43.94$, respectively. The average values of jet power of radiation between FSRQs and BL Lacs are significantly different using the parameter T-test ($P=0.0005$). According to Kolmogorov–Smirnov test ($P=9.18\times10^{-7}$) and Kruskal-Wallis H Test ($P=0.002$), the distributions of jet power of radiation between FSRQs and BL Lacs are significantly different. The FSRQs have a higher average jet power of radiation than BL Lacs. Our results are consistent with the result of \cite{Foschini2015}.  

The average vlues of jet power of electron of FSRQs and BL Lacs are $\langle\log P_{\rm e, FSRQs}\rangle=45.09$ and $\langle\log P_{\rm e, BL Lacs}\rangle=45.07$, respectively. According to the T-test ($P=0.87$), Kolmogorov–Smirnov test ($P=0.96$), and Kruskal-Wallis H Test ($P=0.96$), we find that the distributions of jet power of electron between FSRQs and BL Lacs are not significantly different.

The average values of jet power of the proton of FSRQs and BL Lacs are $\langle\log P_{\rm p, FSRQs}\rangle=45.68$ and $\langle\log P_{\rm p, BL Lacs}\rangle=45.31$, respectively. There is a significant difference in the average jet power of proton between FSRQs and BL Lacs using the T-test ($P=0.006$). Through a Kolmogorov–Smirnov test ($P=0.01$) and Kruskal-Wallis H Test ($P=0.02$), we also find that there is a significant difference in the distribution of jet power of proton between FSRQs and BL Lacs.

The average values of jet power of the magnetic field of FSRQs and BL Lacs are $\langle\log P_{\rm B, FSRQs}\rangle=45.25$ and $\langle\log P_{\rm B, BL Lacs}\rangle=43.59$, respectively. The average values of jet power of the magnetic field between FSRQs and BL Lacs are significant different based on the T-test ($P=2.92\times10^{-16}$). Through a Kolmogorov–Smirnov test ($P=1.07\times10^{-11}$) and Kruskal-Wallis H Test ($P=1.56\times10^{-13}$), the  distributions of jet power of magnetic field between FSRQs and BL Lacs are significantly different. 

The average values of jet power of kinetic of FSRQs and BL Lacs are $\langle\log P_{\rm jet, FSRQs}\rangle=46.52$ and $\langle\log P_{\rm jet, BL Lacs}\rangle=46.09$, respectively. There is a significant difference in the average jet power of kinetic between FSRQs and BL Lacs using the T-test ($P=1.22\times10^{-5}$). Through a Kolmogorov–Smirnov test ($P=2.45\times10^{-8}$) and Kruskal-Wallis H Test ($P=2.42\times10^{-5}$), we also find that there is a significant difference in the distribution of jet power of kinetic between FSRQs and BL Lacs. The average values of jet power of kinetic of FSRQs is larger than that of BL Lacs \citep{Foschini2015}. 

The average values of accretion disk luminosity in Eddington units of FSRQs and BL Lacs are $\langle\log L_{\rm disk}/L_{\rm Edd}\rangle_{\rm FSRQs}=-0.88$ and $\langle\log L_{\rm disk}/L_{\rm Edd}\rangle_{\rm BL Lacs}=-2.75$, respectively. The T-test shows that there is a significant difference between these two averages ($P=1.81\times10^{-100}$). Through a Kolmogorov–Smirnov test ($P=1.48\times10^{-75}$) and Kruskal-Wallis H Test ($P=2.66\times10^{-56}$), we also find that the distributions of accretion disk luminosity in Eddington units between FSRQs and BL Lacs are significantly different. These results may imply that the accretion modes of FSRQs and BL Lacs are different \citep[e.g.,][]{Ghisellini2010, Sbarrato2012, Sbarrato2014}.    
         
\begin{figure}
	\includegraphics[width=8.0cm,height=12.5cm]{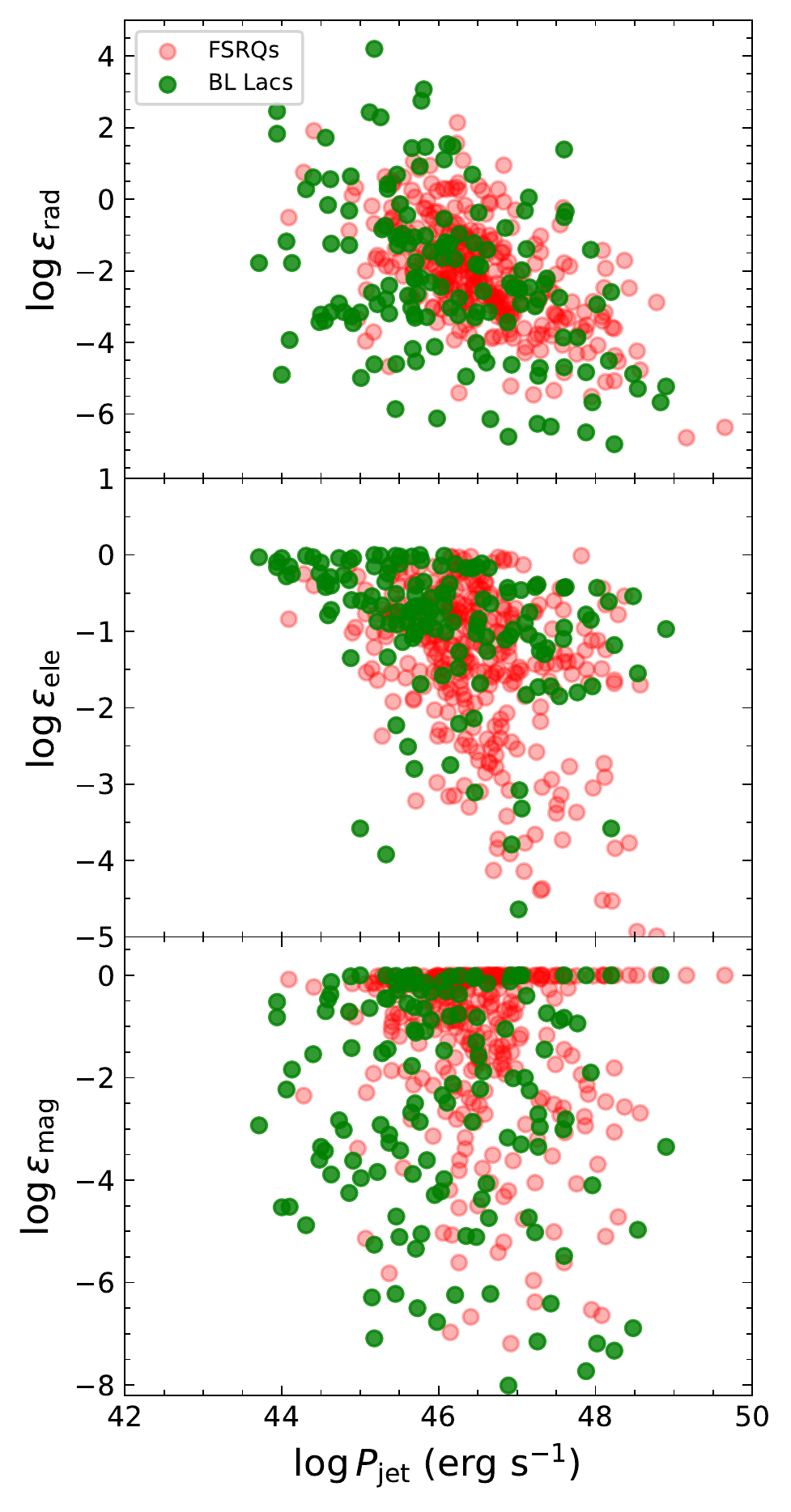}
	\caption{The fraction of the total jet power transformed into radiation (top), relativistic electrons (middle), and Poynting flux (bottom). The red dot is FSRQs and the green dot is BL Lacs.}
	\label{Figure3}
\end{figure}

Figure~\ref{Figure3} shows the fraction of the jet power of kinetic converted to radiation ($\epsilon_{\rm rad}$), carried by relativistic electrons ($\epsilon_{\rm ele}$) and magnetic field ($\epsilon_{\rm mag}$). The average values of $\epsilon_{\rm rad}$ of FSRQs and BL Lacs are $\langle\log \epsilon_{\rm rad, FSRQs}\rangle=-2.04$ and $\langle\log \epsilon_{\rm rad, BL Lacs}\rangle=-2.16$, respectively. The average values of $\epsilon_{\rm ele}$ of FSRQs and BL Lacs are $\langle\log \epsilon_{\rm ele, FSRQs}\rangle=-1.43$ and $\langle\log \epsilon_{\rm ele, BL Lacs}\rangle=-1.03$, respectively. The average values of $\epsilon_{\rm mag}$ of FSRQs and BL Lacs are $\langle\log \epsilon_{\rm mag, FSRQs}\rangle=-1.27$ and $\langle\log \epsilon_{\rm mag, BL Lacs}\rangle=-2.51$, respectively. From the above results, we find that most FSRQs and BL Lacs have $\log\epsilon_{\rm rad}<0$, which implies that the jet power of kinetic of these Fermi blazars is larger than that of the jet power of radiation. \cite{Ghisellini2014} also found that the jet power of kinetic of Fermi blazars is larger than that of the jet power of radiation. Our results are consistent with theirs. At the same time, we also find that almost all of the FSRQs and BL Lacs have $\log\epsilon_{\rm mag}<0$ and hints at a weak magnetization of the emission region, which implies that the jet power of kinetic of these FSRQs and BL Lacs are not dominated by the Poynting flux \citep{Zdziarski2015, Paliya2017}.   

\subsection{The jet formation of Fermi blazars}
\begin{figure}
	\includegraphics[width=8.0cm,height=8.0cm]{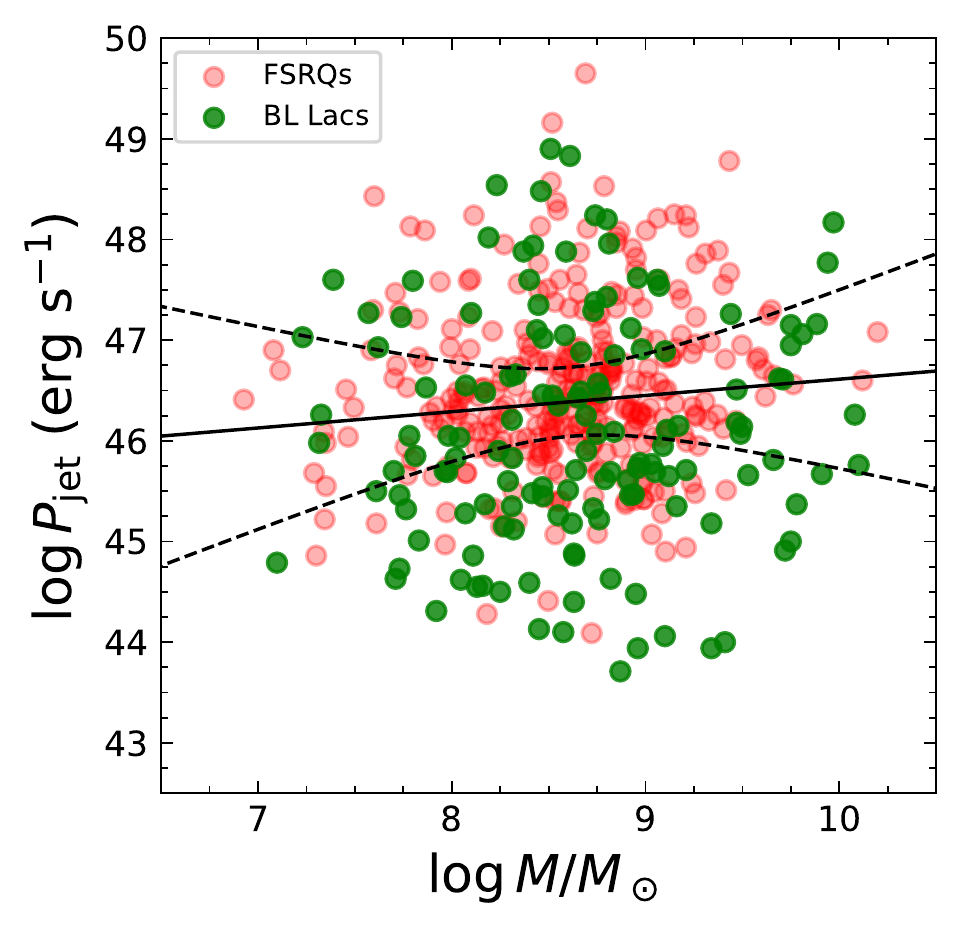}
	\caption{Relation between jet power of kinetic and black hole mass. The red dot is FSRQs and the green dot is BL Lacs. The meaning of solid and dashed lines is the same as that of figure 4. $\log P_{\rm jet}=(0.16\pm0.07)\log M_{\rm BH}+(44.99\pm0.66)$.}
	\label{Figure4}
\end{figure}

The relation between jet power of kinetic and black hole mass for the whole sample is shown in Figure~\ref{Figure4}. We find that there is a moderately strong correlation between jet power of kinetic and black hole mass for the whole sample ($r=0.10, P=0.03$). We also use correlation analysis for every single type of sample. There is a weak correlation between jet power of kinetic and black hole mass for BL Lacs ($r=0.07, P=0.38$). There is a moderately strong correlation between jet power of kinetic and black hole mass for FSRQs ($r=0.13, P=0.02$). Some authors also found a significant relationship between jet power and black hole mass for FSRQs \citep[e.g.,][]{Zhang2014, Xiong2014, Xiao2022}. \cite{Zhang2012} also found a weak correlation between jet power and black hole mass for BL Lacs. Our results are consistent with theirs. \cite{Ghosh1997} suggested that the jet power depends on the black hole mass for a accretion disk dominated by radiation-pressure. \cite{Foschini2011} and \cite{Chen2015b} suggested that the FSRQs are in the radiation-pressure-dominated regime. However, they suggested that the jet power of the BL Lacs does not depend on the mass of the black hole, but on the accretion rate, which implies that the BL Lacs are in the gas-pressure-dominated regime \citep{Foschini2011, Chen2015b}. These results show that FSRQs and BL Lacs have different accretion modes.  
	      
There is evidence that there is a close relationship between jet and accretion in jetted AGNs \citep[e.g.,][]{Rawlings1991, Falcke1995, Cao1999, Wang2004, Liu2006,  Gu2009, Ghisellini2009, Ghisellini2010, Ghisellini2011, Sbarrato2012, Sbarrato2014, Ghisellini2014, Xiong2014, Chen2015a, Zhang2015, Paliya2017, Paliya2019, Xiao2022, Zhang2022, Chen2022}. \cite{Ghisellini2014} used the one-zone lepton model to fit the multi-band data of 217 Fermi blazars to obtain the jet power. They found that there is a strong correlation between jet power and accretion disk luminosity for 217 Fermi blazars. In this work, we use a larger sample of 459 Fermi blazars to restudy the relationship between jet power and accretion disk luminosity. The jet power of 459 Fermi blazars is estimated through the one-zone leptonic model.

\begin{figure}
	\includegraphics[width=8.0cm,height=13.0cm]{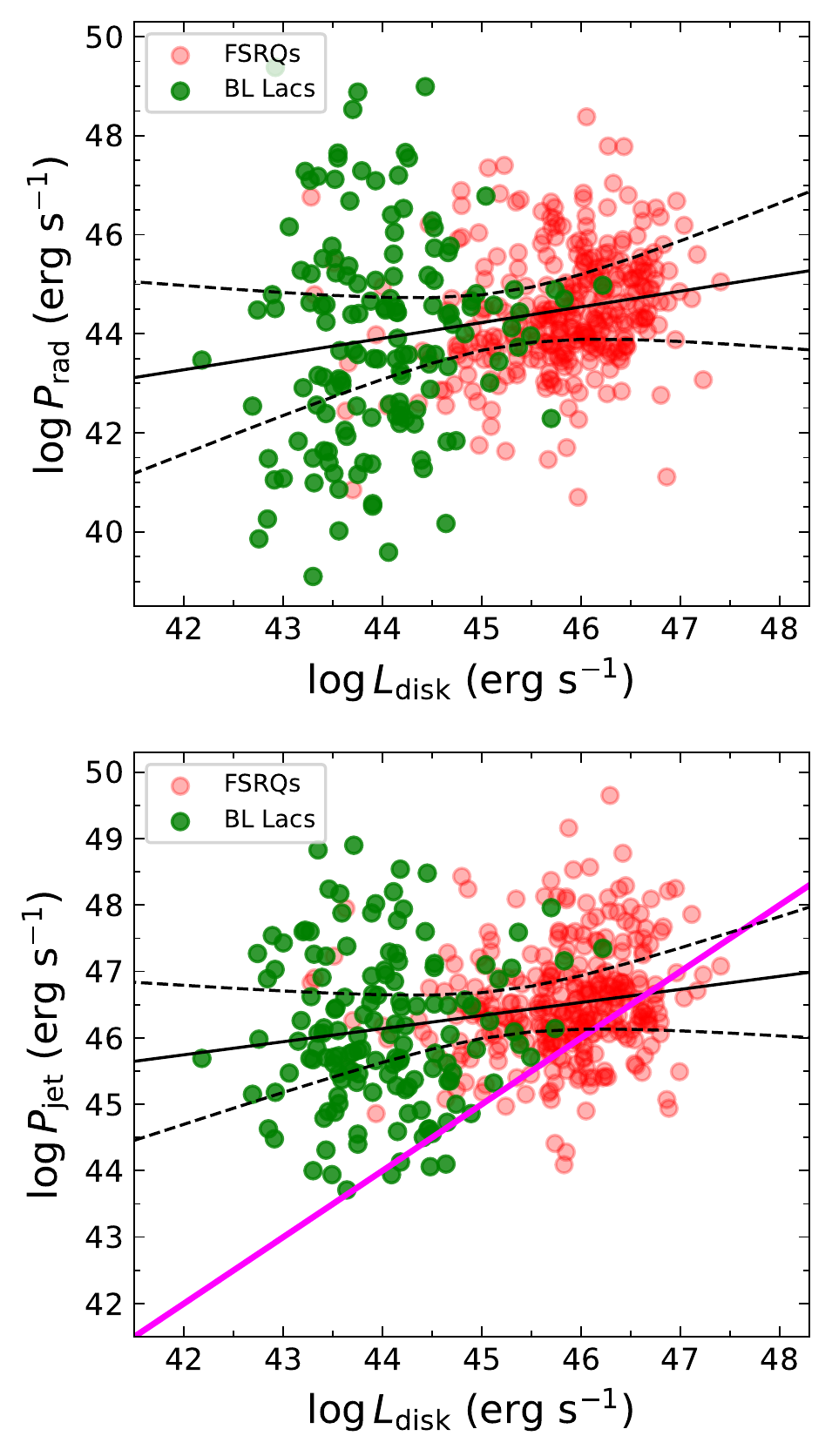}
	\caption{Relation between jet power of radiation (top) and jet power of kinetic (bottom) and accretion disk luminosity for Fermi blazars, respectively. The red dot is FSRQs and the green dot is BL Lacs. The meaning of solid and dashed lines is the same as that of figure 4. In the bottom plots, the pink solid line represents the one-to-one correlation. $\log P_{\rm rad}=(0.31\pm0.06)\log L_{\rm disk}+(29.95\pm2.91)$; $\log P_{\rm jet}=(0.19\pm0.04)\log L_{\rm disk}+(37.46\pm1.79)$.}
	\label{Figure5}
\end{figure}

In the top panel of Figure~\ref{Figure5}, we show the relationship between the jet power of radiation and accretion disk luminosity for all Fermi blazars (FSRQs+BL Lacs). We find a significant correlation between the jet power of radiation and accretion disk luminosity for all Fermi blazars ($r=0.22, P=1.11\times10^{-6}$). The test of Spearman ($r=0.24, P=1.15\times10^{-7}$) and Kendall tau ($r=0.19, P=1.26\times10^{-8}$) also show a significant correlation between jet power of radiation and accretion disk luminosity for all Fermi blazars. However, because jet power of radiation and accretion disk luminosity may depend on redshift, we also perform a partial correlation test. Even after excluding the general redshift dependence, we find that there is always a correlation between the jet power of radiation and the accretion disk luminosity, although the significance becomes a little weak ($r_{par}=0.11, P=0.02$). \cite{Paliya2017} also found that the correlation between the jet power of radiation and the accretion disk luminosity becomes a little weak by using 324 Fermi blazars.    

The relation between jet power of kinetic and accretion disk luminosity for all Fermi blazars is shown in the bottom panel of Figure~\ref{Figure5}. We find a significant correlation between jet power of kinetic and accretion disk luminosity for all Fermi blazars ($r=0.23, P=8.86\times10^{-7}$). The test of Spearman ($r=0.24, P=1.07\times10^{-7}$) and Kendall tau ($r=0.17, P=4.79\times10^{-8}$) also indicate a strong correlation between jet kinetic power and accretion disk luminosity for all Fermi blazars. There is still a moderately strong correlation between jet kinetic power and accretion disk luminosity when redshift is excluded ($r=0.16, P=0.0004$). \cite{Paliya2017} studied the relation between jet kinetic power and accretion disk luminosity for 324 Fermi blazars. They also found a significant correlation between jet power and accretion disk luminosity when redshift is excluded. Our results are consistent with theirs. At the same time, we also find that the jet power of kinetic is slightly larger than the accretion disk luminosity for most Fermi blazars. This is not a coincidence, but the catalytic effect of the magnetic field amplified by the disk. When the magnetic energy density exceeds the energy density of the accretion material near the last stable orbit, the accretion stops and the magnetic energy decreases \citep{Ghisellini2014}. \cite{Ghisellini2014} also suggested that the jet power of kinetic is larger than the accretion disk luminosity for 217 Fermi blazars.                 

\begin{figure}
	\includegraphics[width=8.0cm,height=8.0cm]{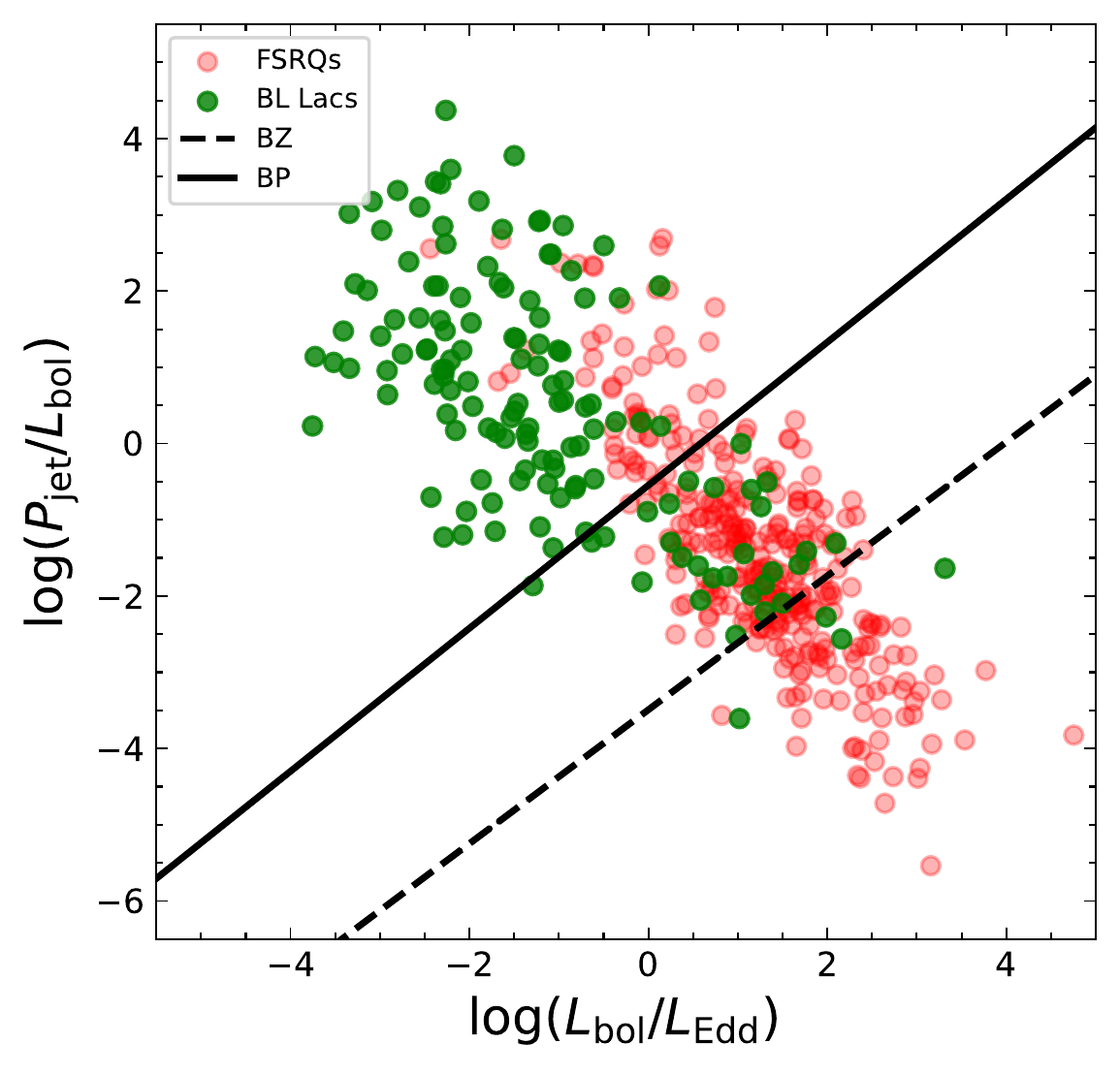}
	\caption{Relation between $L_{\rm bol}/L_{\rm Edd}$ and $P_{\rm jet}/L_{\rm bol}$. The red dot is FSRQs and the green dot is BL Lacs. Solid line: Maximal jet power $P_{\rm BP}^{\rm max}$ extracted from a standard accretion disk (the Blandford-Payne mechanism). Dashed line: Maximal jet power $P_{\rm BZ}^{\rm max}$ extracted from a rapidly spinning black hole $a=0.95$ (the Blandford–Znajek mechanism).}
	\label{Figure6}
\end{figure}

Figure~\ref{Figure6} shows the relation between the ratios $L_{\rm bol}/L_{\rm Edd}$ and $P_{\rm jet}/L_{\rm bol}$. \cite{Nemmen2012} estimated the bolometric
luminosity of Fermi blazars as $L_{\rm bol}=L_{\gamma}+L_{\rm sy}$ where $L_{\gamma}$ is $\gamma$-ray luminosity and $L_{\rm sy}$ is synchrotron peak frequency luminosity. Following the method of \cite{Nemmen2012}, we also use the above formula to get the bolometric luminosity of Fermi blazars. The accretion rate ($\dot{m}=\dot{\rm M}/\dot{M_{\rm Edd}}\simeq L_{\rm bol}/L_{\rm Edd}$) is then estimated for our sample \citep{Cao2004}. We use Equations (3), (4), and (7) to estimate the maximal jet power of the BP mechanism. Similarly, We also use Equations (4), (7), and (10) to estimate the maximal jet power of the BZ mechanism. Because the power of the jet is proportional to the mass of the black hole ($P_{\rm jet} \propto m$, see \cite{Cao2021}), the bolometric luminosity is also proportional to the mass of the black hole ($L_{\rm bol}\propto m$, see \cite{Wu2008}). Thus, the ratio of $P_{\rm jet}/L_{\rm bol}$ is only a function of the spin of black hole and accretion rate ($\dot{m}$) (see details in \cite{Cao2003}). Therefore, we only use the black hole spin $a=0.95$, and the accretion rate $\dot{m}$ is used as a parameter of free variation. We find that the jet kinetic power of about 68\% FSRQs is above the maximal jet power expected to be extracted from the BZ mechanism (Fig.~\ref{Figure6}, dashed line). The jet kinetic power of about 97\% BL Lacs is above the maximal jet power expected to be extracted from the BZ mechanism. These results may indicate that the jet of the Fermi blazars can not be fully explained by the BZ mechanism. \cite{Foschini2011} also found that the BZ mechanism fails to completely account for the jet power of FSRQs. \cite{Chen2015b} compared the maximum jet power of the BZ mechanism with the observed jet power and found that the jet power of Fermi blazars cannot be fully explained by the BZ mechanism (see Figure 4 of \cite{Chen2015b}). 

When considering that the maximum jet power is expected to be extracted from the a magnetized accretion disk, we find that the jet kinetic power of about 83\% FSRQs is below the maximal jet power expected to be extracted from a magnetized accretion disk (Fig.~\ref{Figure6}, solid line). The jet kinetic power of about 23\% BL Lacs is below the maximal jet power expected to be extracted from a magnetized accretion disk. These results may suggest that the jets of FSRQs are mainly generated by the BP mechanism. \cite{Paliya2021} concluded that the overall physical properties of Fermi blazar are likely to be controlled by the accretion rate in the Eddington unit. In particular, FSRQs have high accretion rates in the Eddington unit. \cite{Xiao2022} proposed that the jets of FSRQs are powered mostly by the accretion disk. The jets of the remaining 27\% FSRQs may need to be explained by other jet models, such as magnetization-driven outflows. \cite{Cao2018} suggested that the magnetic field dragged inward by the accretion disk with magnetic outflows may accelerate the jets in blazars. 

The jet kinetic power of most BL Lacs can not be explained by both the BZ mechanism and BP mechanism when considering the standard thin disk. \cite{Cao2003} also found that the jet power of BL Lacs can not be explained by both the BZ mechanism and BP mechanism when considering the standard thin disk by using 29 BL Lacs (Figure 1 of \cite{Cao2003}). We confirm the results of \cite{Cao2003}. We find that most BL Lacs in our sample have low accretion rates. The source with advection-dominated accretion flows (ADAFs) usually has a low accretion rate \citep[e.g.,][]{Narayan1995}. These results may imply that its accretion disk is not a standard thin disk but an ADAFs. \cite{Cavaliere2002} have proposed that ADAFs might be present in most BL Lacs. \cite{Cao2002} have suggested that most BL Lacs may have ADAFs surrounding their massive black holes.      

\subsection{The jet kinetic power versus $\gamma$-ray luminosity and radio luminosity}
\begin{figure}
	\includegraphics[width=8.0cm,height=13.0cm]{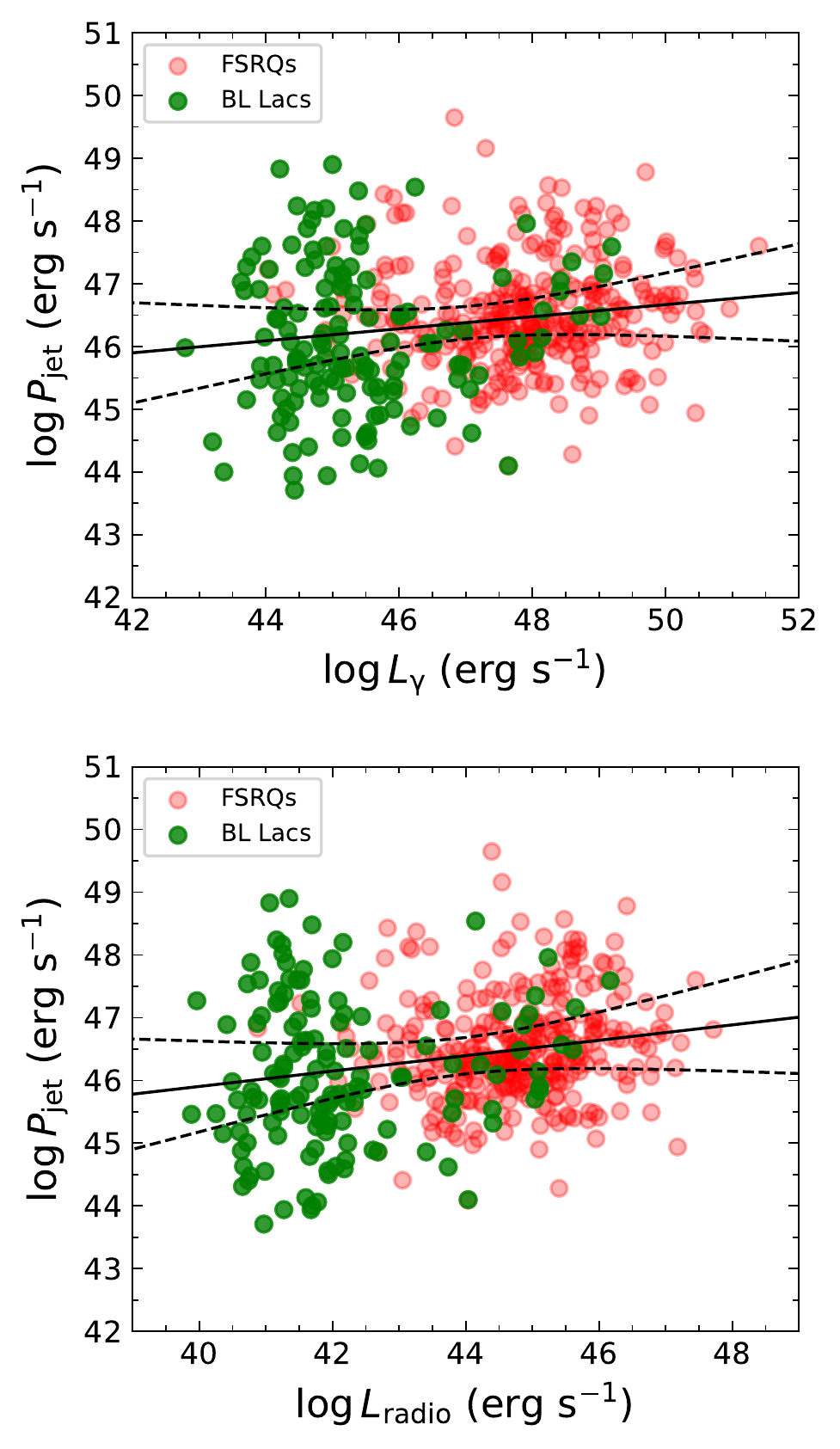}
	\caption{Relation between gamma-ray luminosity (top) and radio luminosity (bottom) and jet kinetic power for Fermi blazars, respectively. The red dot is FSRQs and the green dot is BL Lacs. The meaning of solid and dashed lines is the same as that of figure 4.}
	\label{Figure7}
\end{figure}

Figure~\ref{Figure7} shows jet power of kinetic as a function of $\gamma$-ray luminosity (top) and 1.4GHz radio luminosity (bottom). We find a strong correlation between jet power of kinetic and $\gamma$-ray luminosity for Fermi blazars ($r=0.17, P=0.0001$). The regression result gives 

\begin{equation}
\log P_{\rm jet}=(0.10\pm0.03)\log L_{\gamma}+(41.86\pm1.20). 
\end{equation}

There is also a strong correlation between jet power of kinetic and 1.4GHz radio luminosity for Fermi blazars ($r=0.22, P=1.77\times10^{-6}$). The regression result gives  

\begin{equation}
	\log P_{\rm jet}=(0.12\pm0.03)\log L_{\rm radio}+(40.99\pm1.12)
\end{equation}
These results may imply that gamma-ray and radio emissions originated from the jet. The gamma-ray luminosity and radio luminosity can be used to indicate the jet power of kinetic of Fermi blazars. Many authors have confirmed that there is a significant correlation between jet power and radio luminosity by using small samples \citep[e.g.,][]{Willott1999, Birzan2008, Cavagnolo2010}. \cite{Nemmen2012} also found a significant correlation between jet power and gamma-ray luminosity by using 234 Fermi blazars.  

\section{Conclusions}
With the release of the fourth source catalog data (4FGL-DR2) of the Fermi telescope \citep{Abdollahi2020}, we can obtain high-quality quasi-simultaneous multiwavelength data of a large sample of Fermi blazar. This enables us to use a simple one-zone leptonic emission model to fit the quasi-simultaneous multiwavelength data of a larger sample of Fermi blazars and obtain some jet physical parameters, such as magnetic field, Doppler factor, jet power, and so on. At the same time, compared with the theoretical model of jets, we further discussed the jet formation mechanism of the Fermi blazar. The main results are as follows:

(1) Compared with BL Lacs, FSRQs have a higher average magnetic field, jet power of radiation, jet power of proton, jet power of the magnetic field, jet kinetic power, and accretion disk luminosity in Eddington units. According to a parameter T-test, nonparametric Kolmogorov–Smirnov test, and Kruskal-Wallis H Test, we find that the distributions of these physical parameters between FSRQs and BL Lacs are significantly different. However, there is no significant difference between FSRQs and BL Lacs in the distribution of the Doppler factor and jet power of electrons.  
 
(2) The Fermi blazars have $\log\epsilon_{\rm rad}<0$, which implies that the jet power of kinetic of these Fermi blazars is larger than that of the jet power of radiation. At the same time, we also find that almost all of the FSRQs and BL Lacs have $\log\epsilon_{\rm mag}<0$, which implies that the jet power of kinetic of these FSRQs and BL Lacs are not dominated by the Poynting flux.

(3) There is a weak correlation between jet power of kinetic and black hole mass for Fermi blazars. However, there is a moderately strong correlation between jet power of kinetic and black hole mass for FSRQs.

(4) Even if the redshift is excluded, there has always been a strong correlation between jet power and accretion disk luminosity for Fermi blazars, indicating a close relationship between jet and accretion. 

(5) We find that the jet power of kinetic of about 82\% FSRQs is below the maximal jet power expected to be extracted from a magnetized accretion disk. This result may imply that the jet of FSRQs is mainly generated by the BP mechanism. However, the jet of BL Lacs can not be explained by both the BZ mechanism and BP mechanism in the case of a standard thin disk. At the same time, the BL Lacs have low accretion rates. These results may imply that the BL Lacs have ADAFs surrounding their massive black holes.

(6) There is a significant correlation between jet power of kinetic and gamma-ray luminosity and 1.4GHz radio luminosity for Fermi blazars. This result suggests that the jet dominates the emission of gamma-ray and radio. The gamma-ray luminosity and radio luminosity can be used to indicate the jet kinetic power of Fermi blazars.

\acknowledgments We are very grateful to the referee for the very helpful
comments and suggestions. The work is supported by the National Natural Science Foundation of China (No. 12203028, 11733002, 12121003, 12192220, 12192222 and U2031201). This work was support from the research project of Qujing Normal University (2105098001/094). This work is supported by the youth project of Yunnan Provincial Science and Technology Department (202101AU070146, 2103010006). Yongyun Chen is grateful for funding for the training Program for talents in Xingdian, Yunnan Province. We also acknowledge the science research grants from the China Manned Space Project with NO. CMS-CSST-2021-A05. Thanks to Cao Xinwu for his suggestions and opinions.

\bibliographystyle{aasjournal}
\bibliography{example}    

\label{lastpage}

\end{document}